\documentclass[a4paper]{article}
\usepackage{graphicx}
\usepackage{subfig}
\usepackage{cite}
\title{Invariants of Braided Ribbon Networks}
\author{Jonathan Hackett}

\begin{document}

\maketitle

\begin{abstract}
We present a consistent definition for braided ribbon networks in 3-dimensional manifolds, unifying both three and four valent networks in a single framework. We present evolution moves for these networks which are dual to the Pachner moves on simplices and present an invariant of this evolution. Finally we relate these results back to previous work in the subject.
\end{abstract}

In \cite{BilsonThompson:2005bz} a proposal was presented for mapping local excitations of a background independent model of quantum space-time to fermions of the standard model. This result led to a research program \cite{BilsonThompson:2006yc, Hackett:2007dx,Wan:2007nf,Smolin:2007sn,Hackett:2008ie, BilsonThompson:2008ex,Hackett:2008tt,He:2008is, He:2008jc, Wan:2008qs, BilsonThompson:2009fh} devoted to studying such models, and investigating whether - as suggested by that early result - they could provide a mechanism for the emergence of particle physics from theories of quantum gravity. These works have all used a common framework referred to as \textit{Braided Ribbon Networks} (BRNs), which are a combination of forms from \cite{Markopoulou:1997hu,Smolin:2002sz}.

Despite the progress that has been made in studying the BRNs, this progress has been made in an ad hoc fashion in either a 3-valent or 4-valent formalisms without significant crossover between the two. This is unfortunate as the progress in each of the valences are complementary to one another, and so both would benefit from a coherent underlying formalism.

We will present a coherent framework for BRNs, and demonstrate how these frameworks coincide with the previous definitions used in the literature. We will then demonstrate that this coherent framework allows the concept of the reduced link from the 3-valent formalism to be extended to 4-valent networks as well.

\section{Braided Ribbon Networks}
\label{BRNS}
We introduce Braided Ribbon Networks of valence $n$ (with $n \geq 3$) as follows:
\begin{quote}
We begin by considering an $n$-valent graph embedded in a compact $3$ dimensional manifold. We construct a $2$-surface from this by replacing each node by a $2$-sphere with $n$ punctures ($1$-sphere boundaries on the $2$-sphere), and each edge by a tube which is then attached to each of the nodes that it connects to by connecting the tube to one of the punctures on the $2$-sphere corresponding to the node.

Lastly we add to each tube $n-1$ curves from one puncture to the other and then continue these curves across the sphere in such a way that each of the $n$ tubes connected to a node shares a curve with each of the other tubes.

We will freely call the tubes between spheres \textit{edges}, the spheres \textit{nodes} and the curves on the tubes \textit{racing stripes} (terminology from \cite{Smolin:2002sz, Crane:1991ke, Crane:1991jv}) or less formally \textit{stripes}.

We will call a Braided Ribbon Network the equivalence class of smooth deformations of such an embedding that do not involve intersections of the edges or the racing stripes.
\end{quote}

We immediately face the following consequence: under this definition there are only braided ribbon networks of valence $2$,$3$ or $4$ (with valence $2$ being a collection of framed loops). To see this fact we consider a $5$ valent node - a $2$-sphere with $5$ punctures, with each puncture is connected to each other puncture by a non-intersecting curve. Taking each puncture as a node, and the curves as edges, we then get that these objects would constitute the complete graph on $5$ nodes and as they lie in the surface of a $2$-sphere, such a graph would have to be planar. This is impossible by Kuratowski's theorem\cite{kk}: the complete graph on $5$ nodes is non-planar. Likewise, we have for any higher valence $n$ that the graph that would be constructed would have the complete graph on $5$ nodes as a subgraph, and so they too can not be planar. If the reader desires an intuition for this, it may be instructive to recall that these statements are equivalent to the four colour theorem - the existence of such a node would imply the existence of a map requiring five (or more) colours.

\begin{figure}[!h]
  \begin{center}
    \includegraphics[scale=0.2]{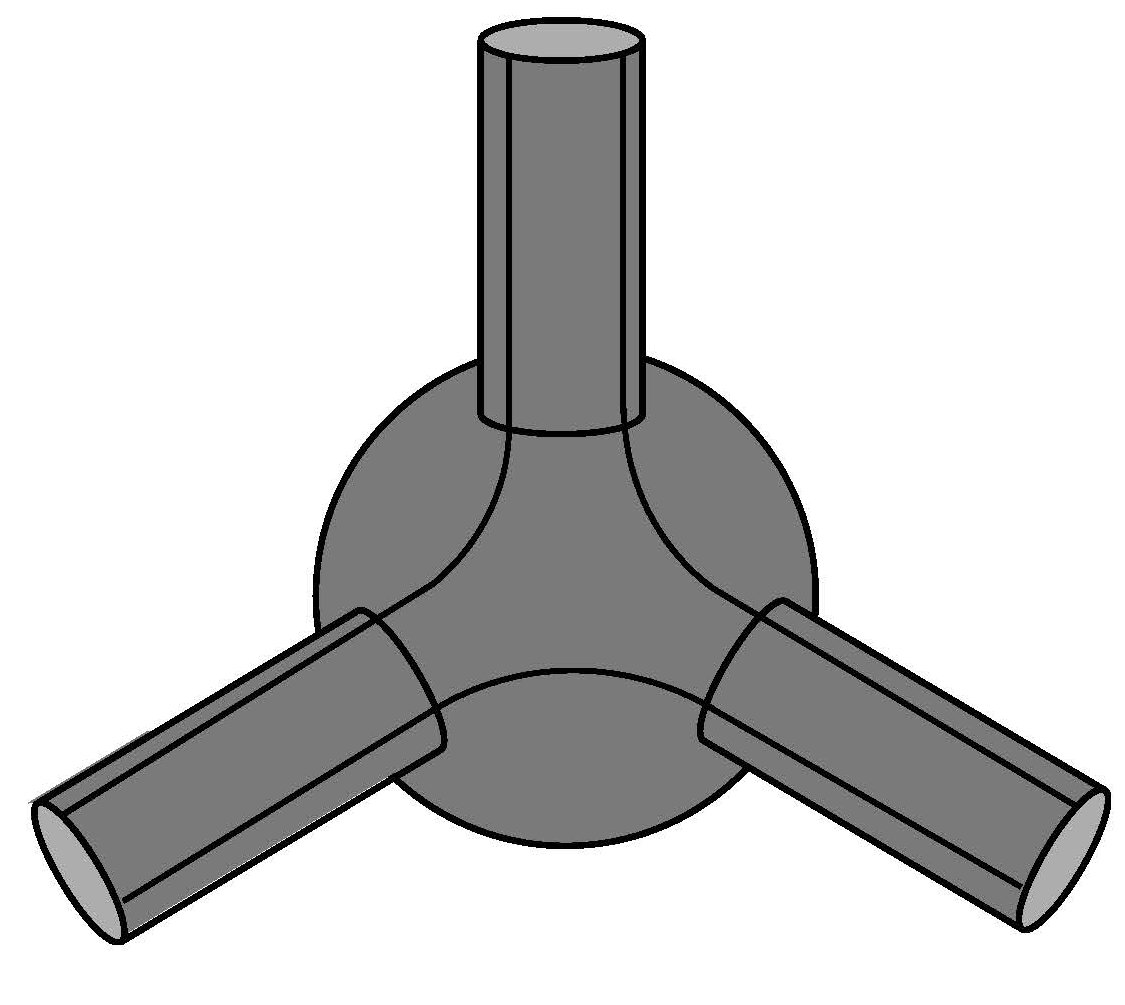}
  \end{center}
\caption{Three Valent Node}
\end{figure}

\begin{figure}[!h]
  \begin{center}
    \includegraphics[scale=0.2]{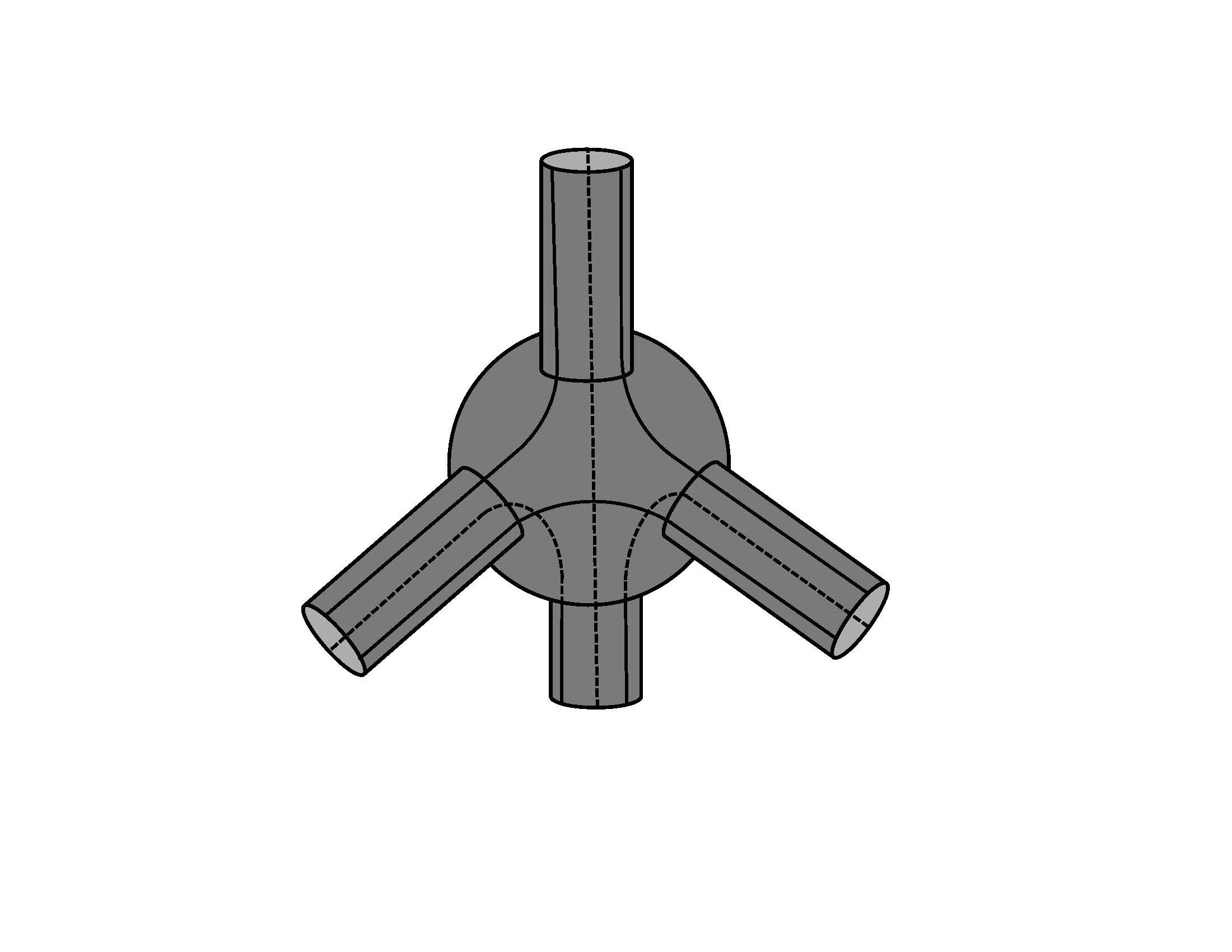}
  \end{center}
\caption{Four Valent Node}
\end{figure}
We will demonstrate later that this restriction is natural, and that it could be lifted if we instead allowed the dimension of the manifold to increase, along with the dimension of the surface which we extend the graph into.

\subsection{Duality of nodes to simplices}
\label{dual}
The nodes of an $n$ valent braided ribbon network can each individually be considered dual to an $n-1$ simplex. The edges which intersect the node are identified with the $n-2$ faces of the simplex, and the curves on the edges are identified with the $n-3$ faces of the simplex. We require that the map from the BRN node to the simplex be consistent in the following way: when a curve is common to two edges that intersect a node, we require that the $n-3$ face that it is identified with is the $n-3$ face common to the $n-2$ faces that the two edges are identified with.

\begin{figure}[!h]
  \begin{center}
    \includegraphics[scale=0.2]{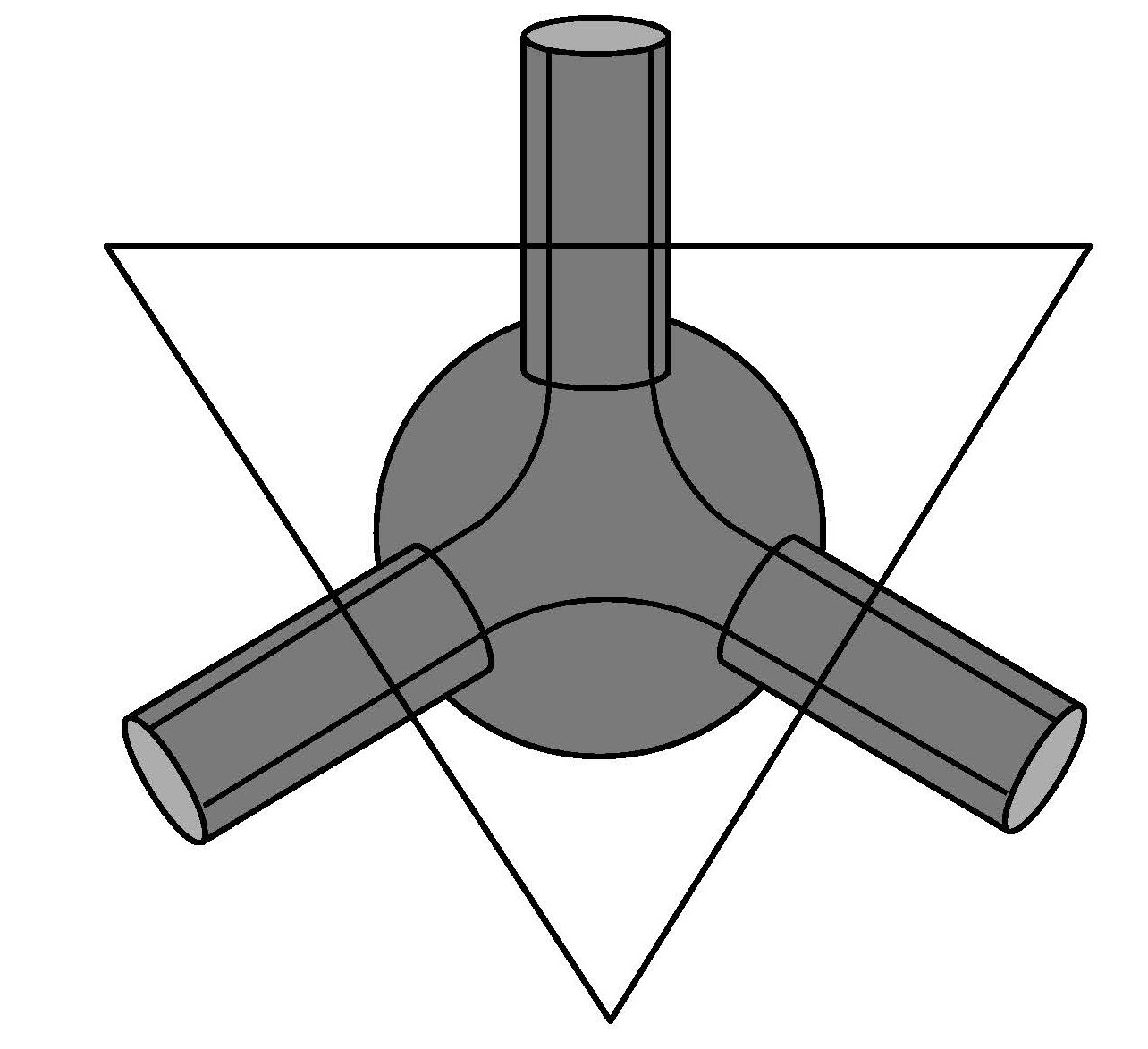}
  \end{center}
\caption{Three valent node dual to a triangle} \label{1}
\end{figure}

\begin{figure}[!h]
  \begin{center}
    \includegraphics[scale=0.2]{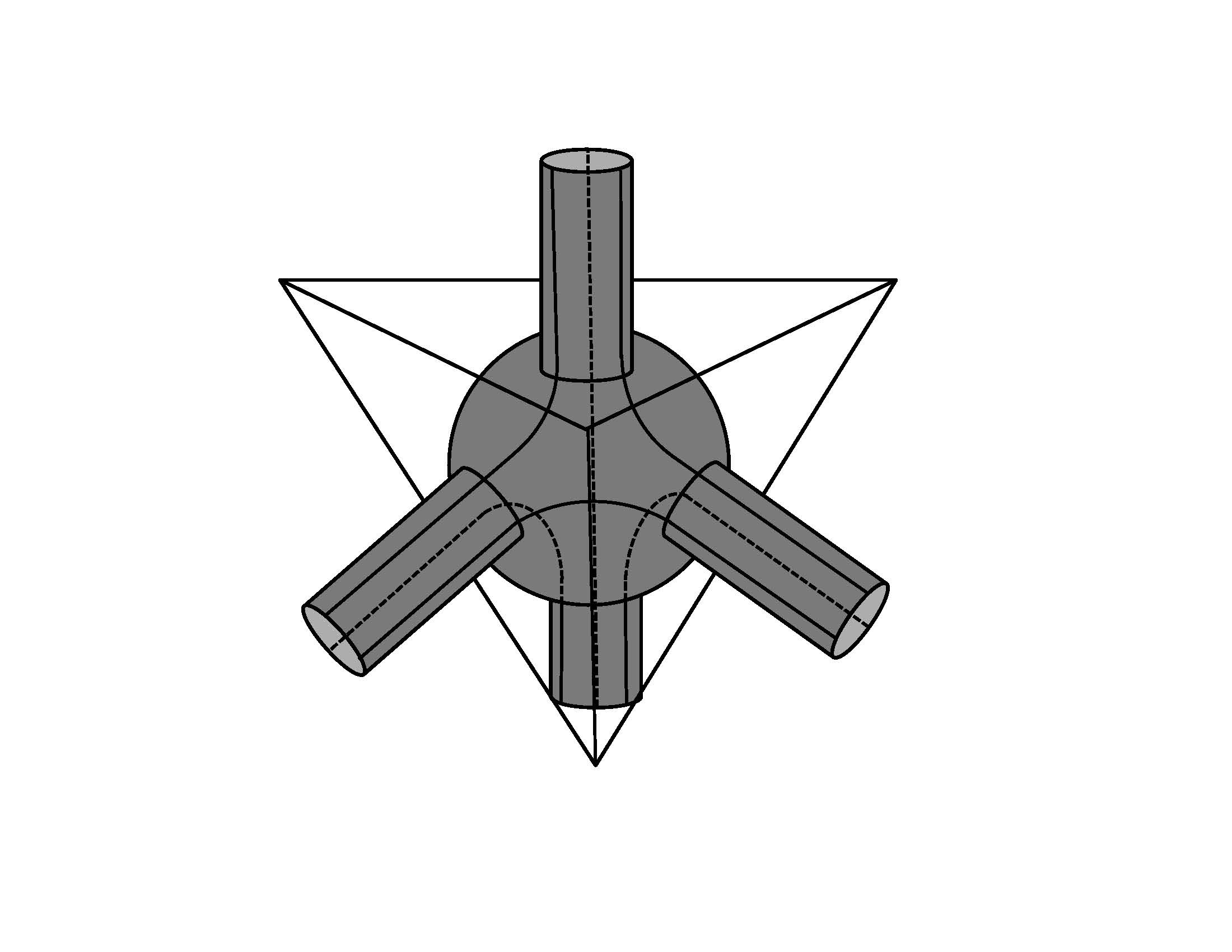}
  \end{center}
\caption{Four valent node dual to a tetrahedron} \label{2}
\end{figure}

For $3$ valent Braided Ribbon Networks this identification gives us that each node is dual to a triangle (see figure \ref{1}), whereas for $4$ valent Braided Ribbon Networks this identification give us a tetrahedron for each node (see figure \ref{2}). This gives some insight into the restriction on the valence of Braided Ribbon Networks: a $5$ valent node would be dual to a $4$-simplex, which given that the graph is embedded in $3$ dimensions would be a significant feat, whereas if we extended to a higher dimensional surface embedded in a higher dimensional manifold, this should be possible.

\begin{figure}[!h]
  \begin{center}
    \includegraphics[scale=0.2]{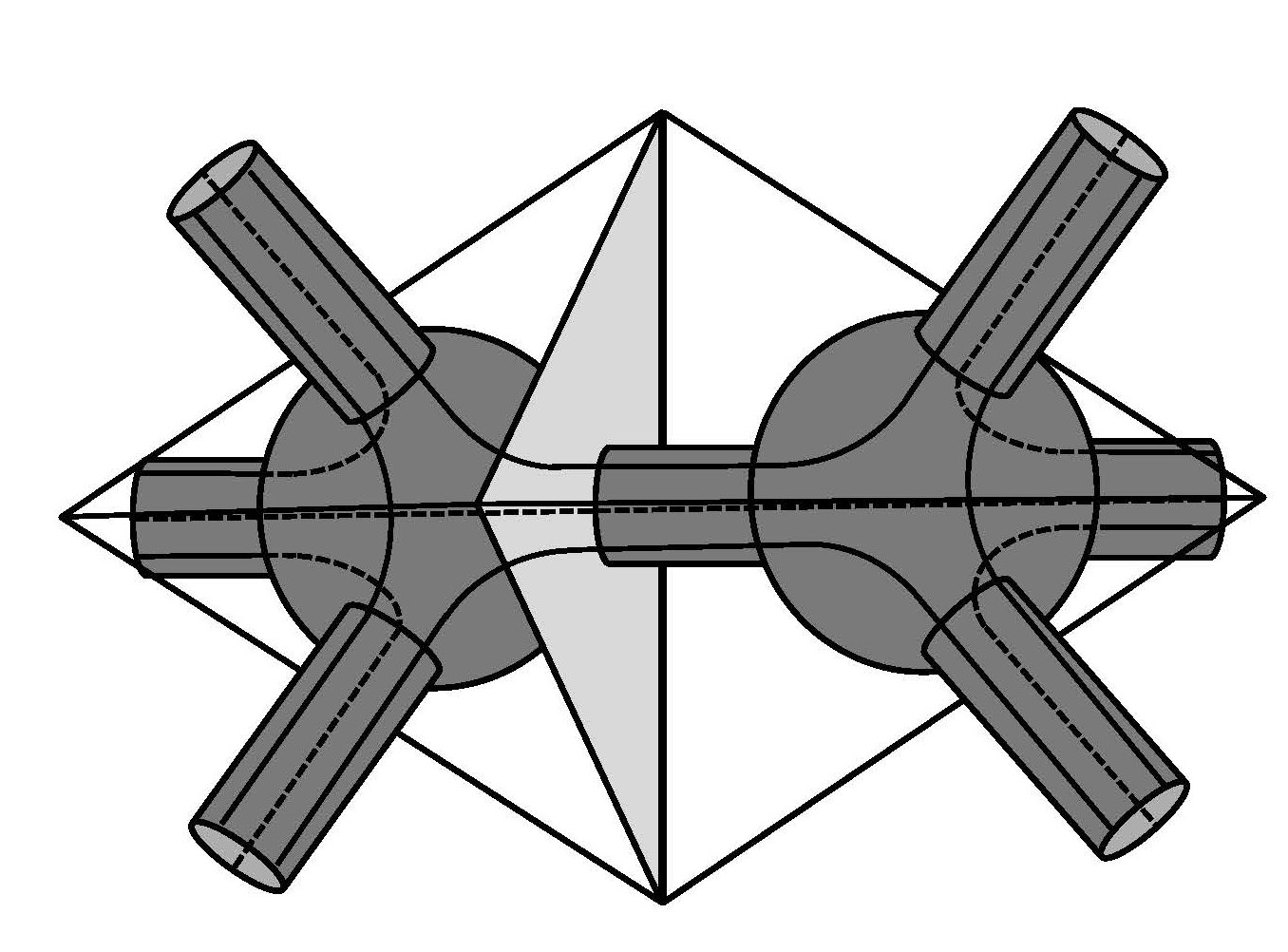}
  \end{center}
\caption{Gluing of two simplices dual to nodes} \label{3}
\end{figure}

We extend this notion of duality to multiple nodes in a significantly restricted manner. If an edge connecting two nodes has neither knotting nor twisting we will call the edge \textit{simple} - we extend the dual picture to gluings of simplices to allow the $n-1$ simplex of each node to be glued together along the $n-2$ corresponding to the shared edge. We continue to require the consistency of all of the shared sub-simplices in this scenario (see figure \ref{3}). We will refer to a collection of nodes such that any edges which connect them are simple as being \textit{simply glued}.

\subsection{The Evolution Moves}
\label{pachner}
With the duality established above we define evolution moves on the graph by making reference to the Pachner moves on the dual gluings of $n-1$ simplices. The operations are defined as preserving the external identifications of sub simplices, and we will go through these explicitly for both $3$ and $4$ valent BRNs.

Pachner moves generate transformations between triangulations of piecewise linear manifolds. They take the form of moves between triangulations which have the same boundary. We label them by the number of glued simplices which the move has as its origin and the number that they have as their target. We will then construct from these operations corresponding operations on collections of simply glued nodes.

\begin{figure}[!h]
  \begin{center}
    \includegraphics[scale=0.2]{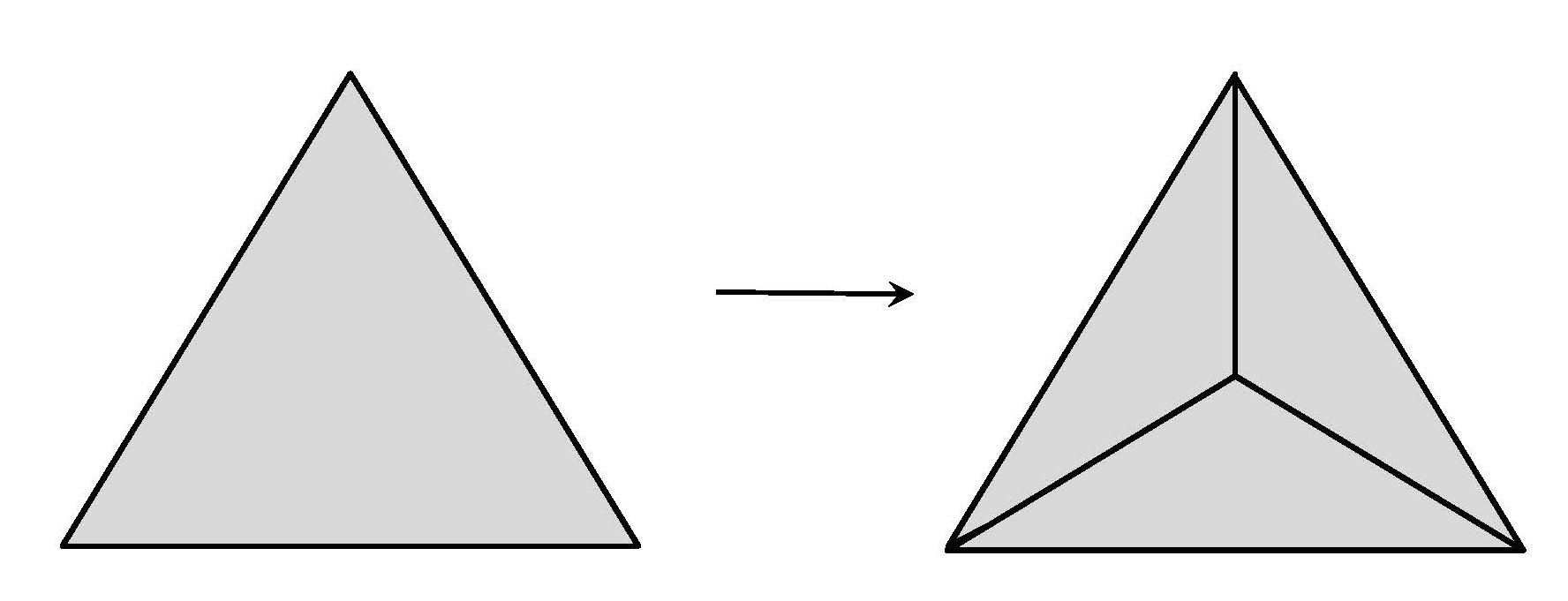}
  \end{center}
\caption{The 1-3 Pachner move} \label{4}
\end{figure}

\begin{figure}[!h]
  \begin{center}
    \includegraphics[scale=0.2]{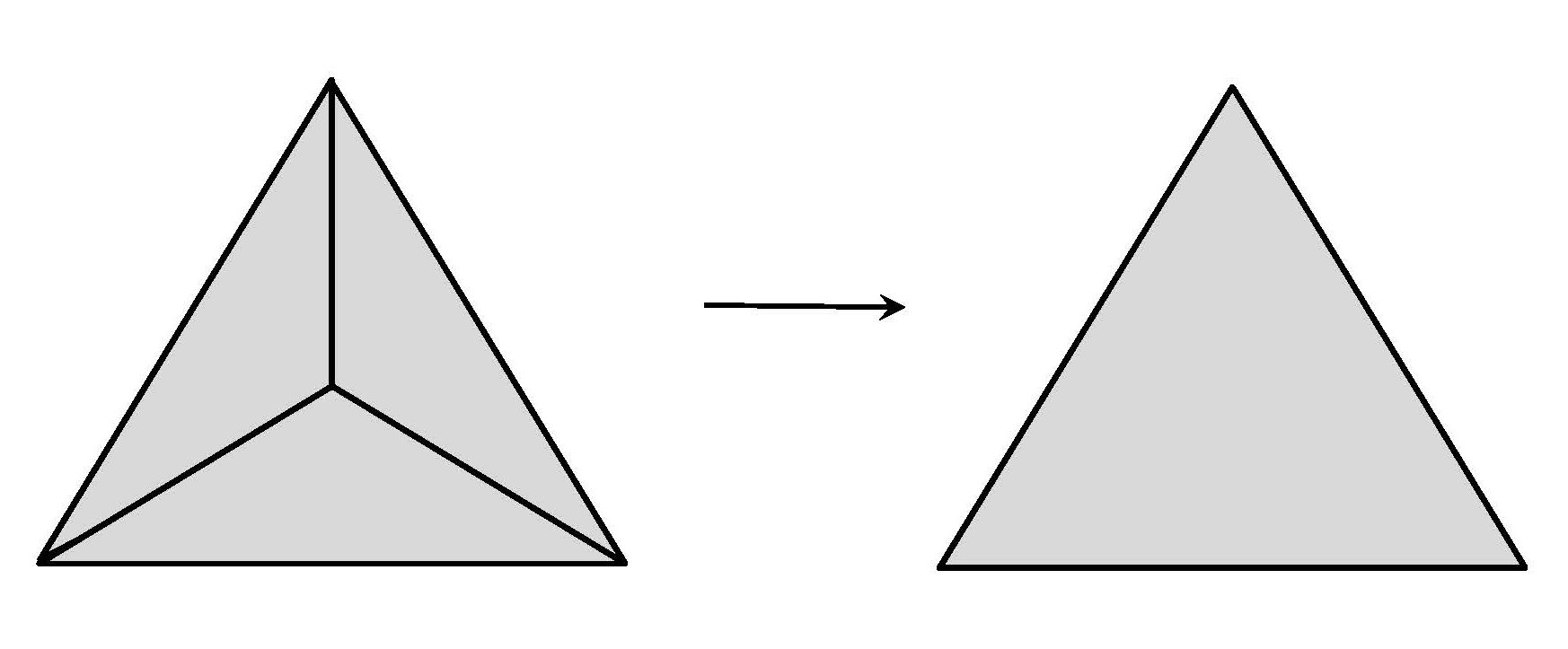}
  \end{center}
\caption{The 3-1 Pachner move} \label{5}
\end{figure}

\begin{figure}[!h]
  \begin{center}
    \includegraphics[scale=0.2]{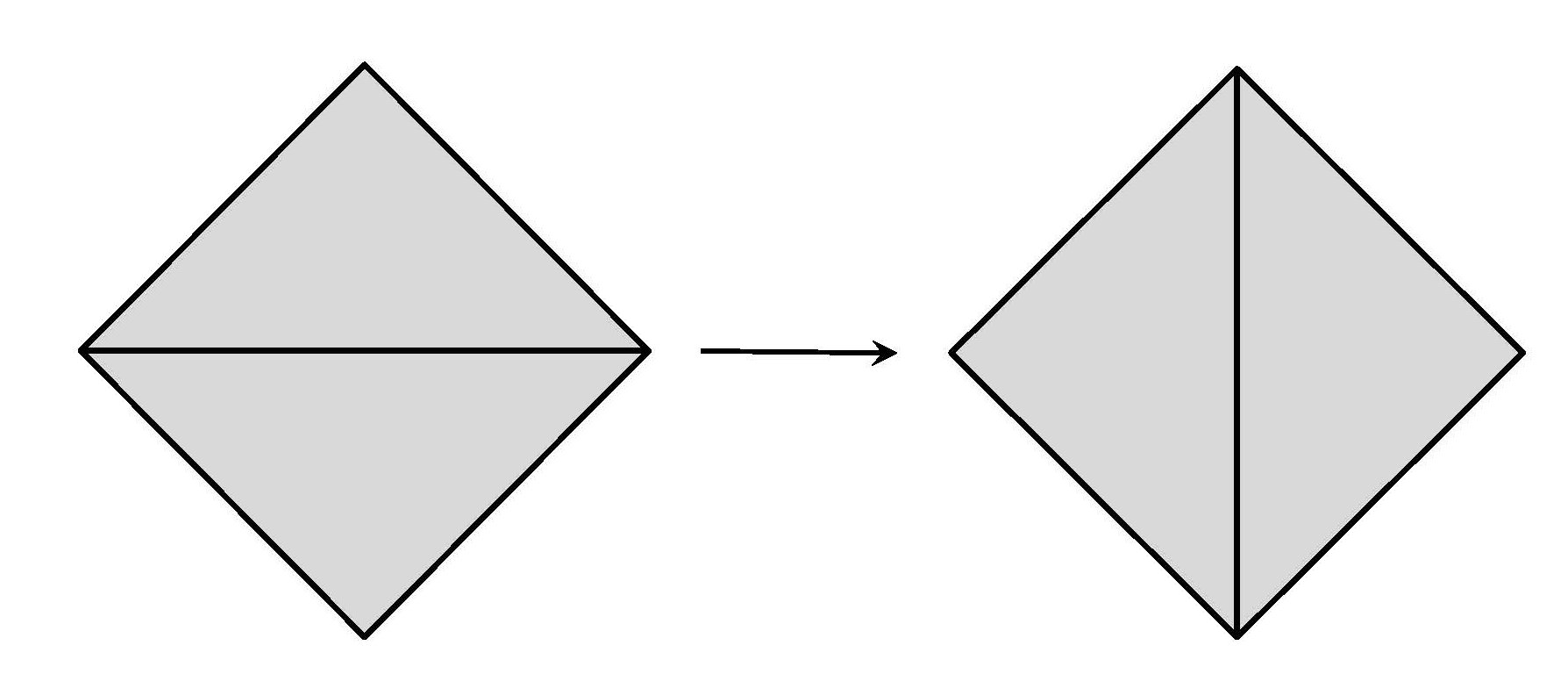}
  \end{center}
\caption{The 2-2 Pachner move} \label{6}
\end{figure}

\begin{figure}[!h]
  \begin{center}
    \includegraphics[scale=0.2]{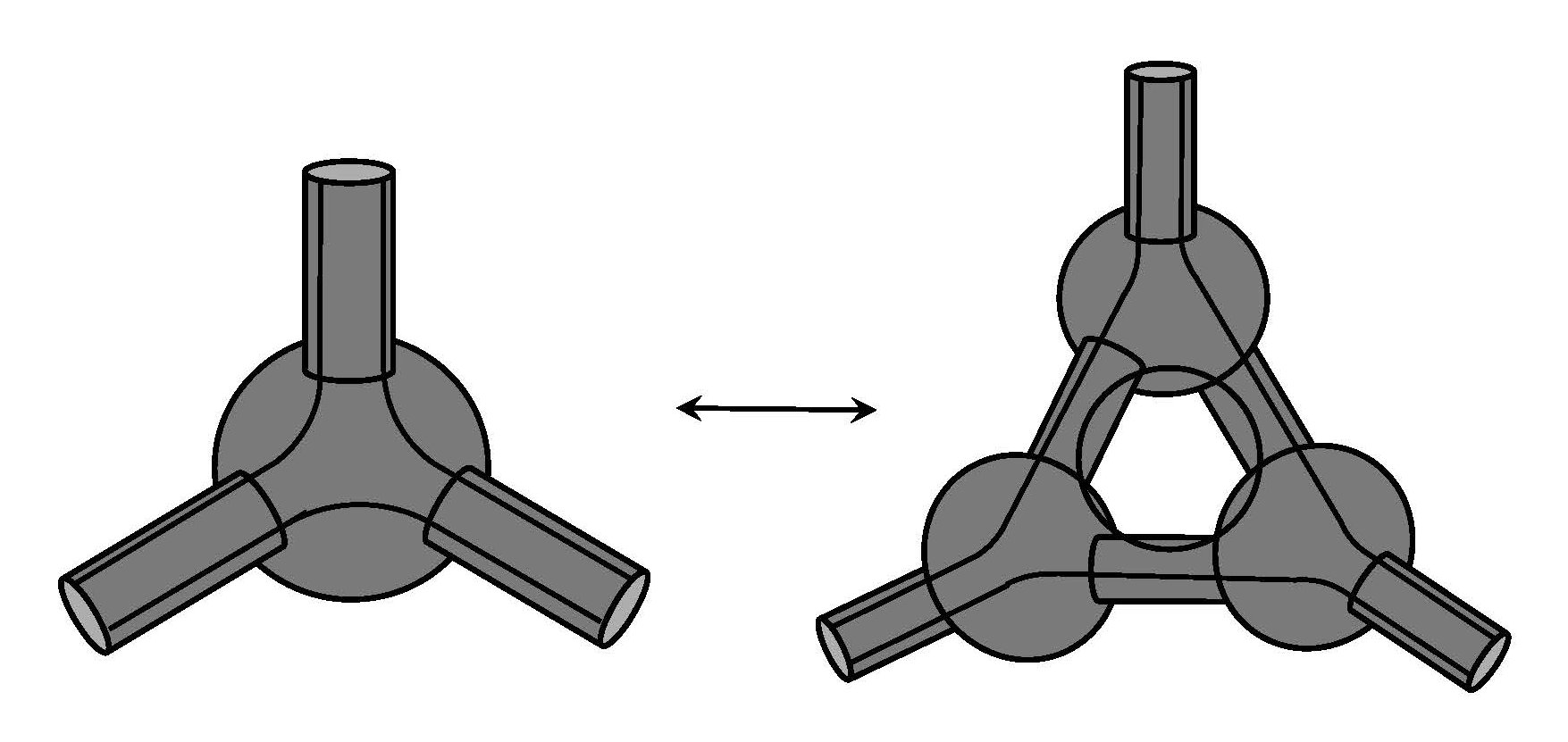}
  \end{center}
\caption{The 1-3 and 3-1 evolution moves} \label{7}
\end{figure}

\begin{figure}[!h]
  \begin{center}
    \includegraphics[scale=0.2]{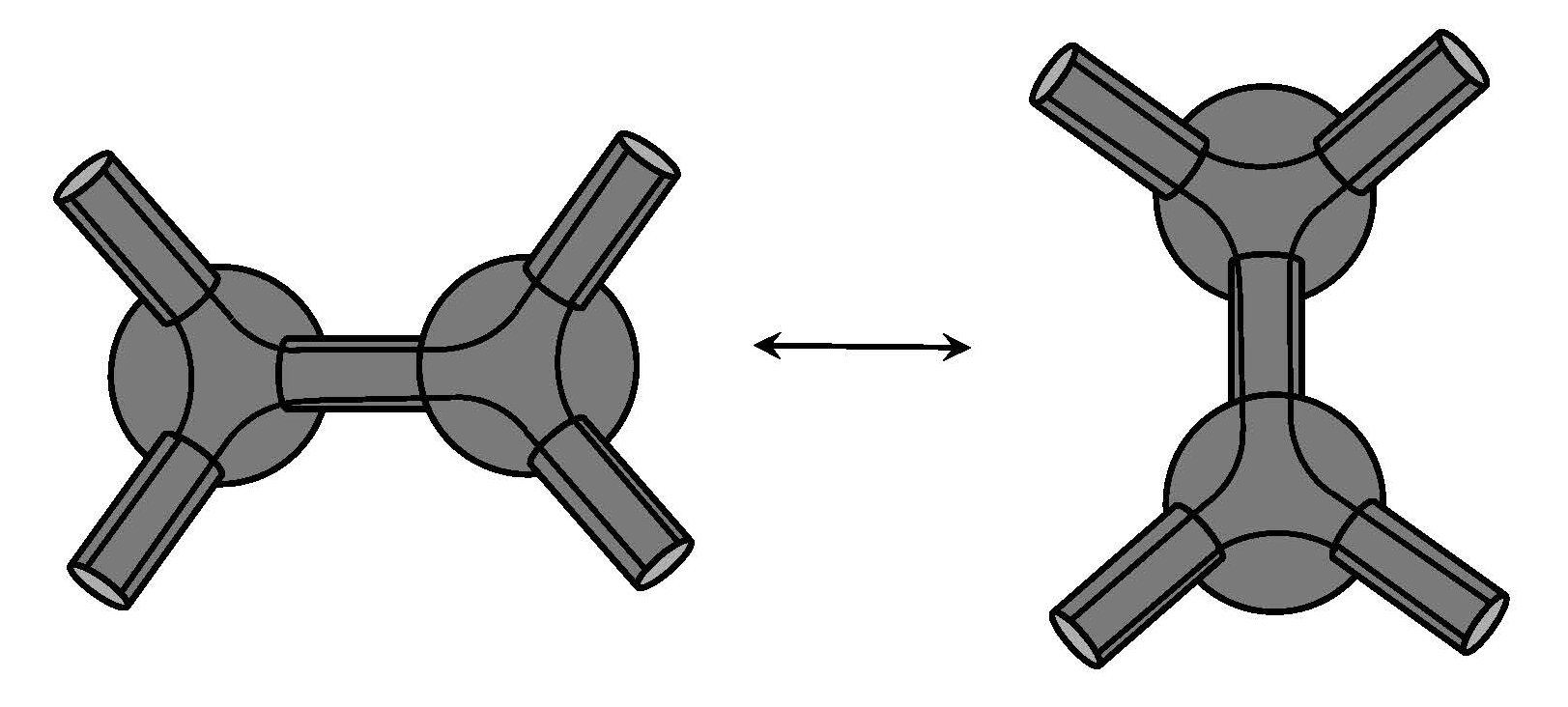}
  \end{center}
\caption{The 2-2 evolution move} \label{8}
\end{figure}

For $3$ valent networks, the corresponding simplices are triangles. The Pachner moves on triangles are the $1-3$ move (figure \ref{4}), the $3-1$ move (figure \ref{5}) and the $2-2$ move (figure \ref{6}). These correspond to similarly named evolution moves for the braided ribbon networks (figures \ref{7} and \ref{8}).

\begin{figure}[!h]
  \begin{center}
    \includegraphics[scale=0.2]{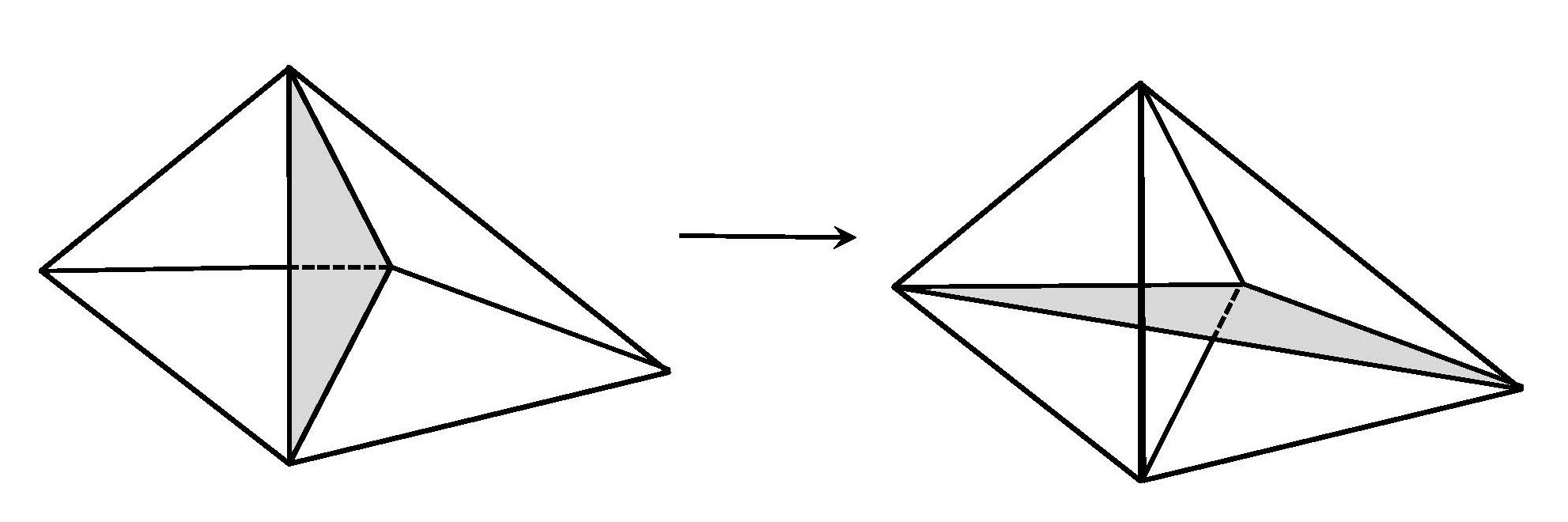}
  \end{center}
\caption{The 2-3 and 3-2 Pachner moves} \label{9}
\end{figure}

\begin{figure}[!h]
  \begin{center}
    \includegraphics[scale=0.2]{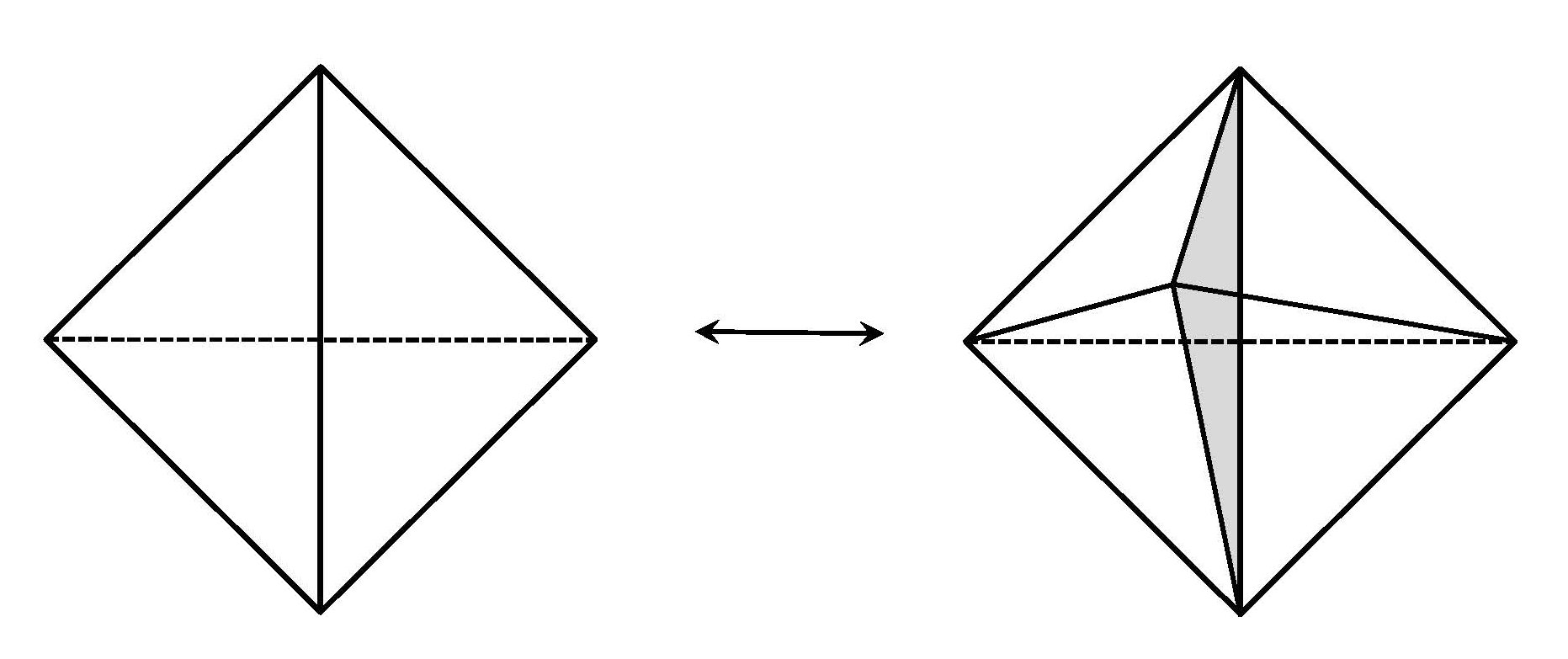}
  \end{center}
\caption{The 1-4 and 4-1 Pachner move} \label{10}
\end{figure}

\begin{figure}[!h]
  \begin{center}
    \includegraphics[scale=0.2]{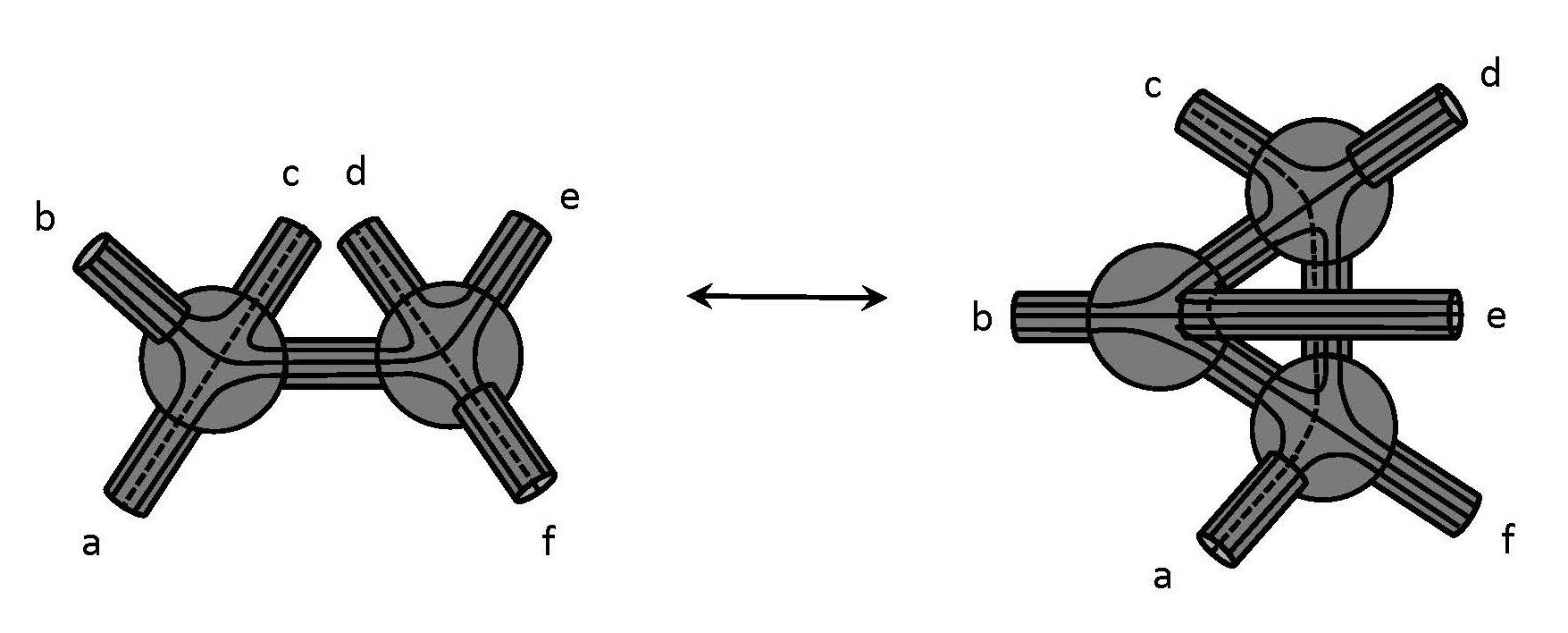}
  \end{center}
\caption{The 2-3 and 3-2 evolution moves} \label{11}
\end{figure}

\begin{figure}[!h]
  \begin{center}
    \includegraphics[scale=0.2]{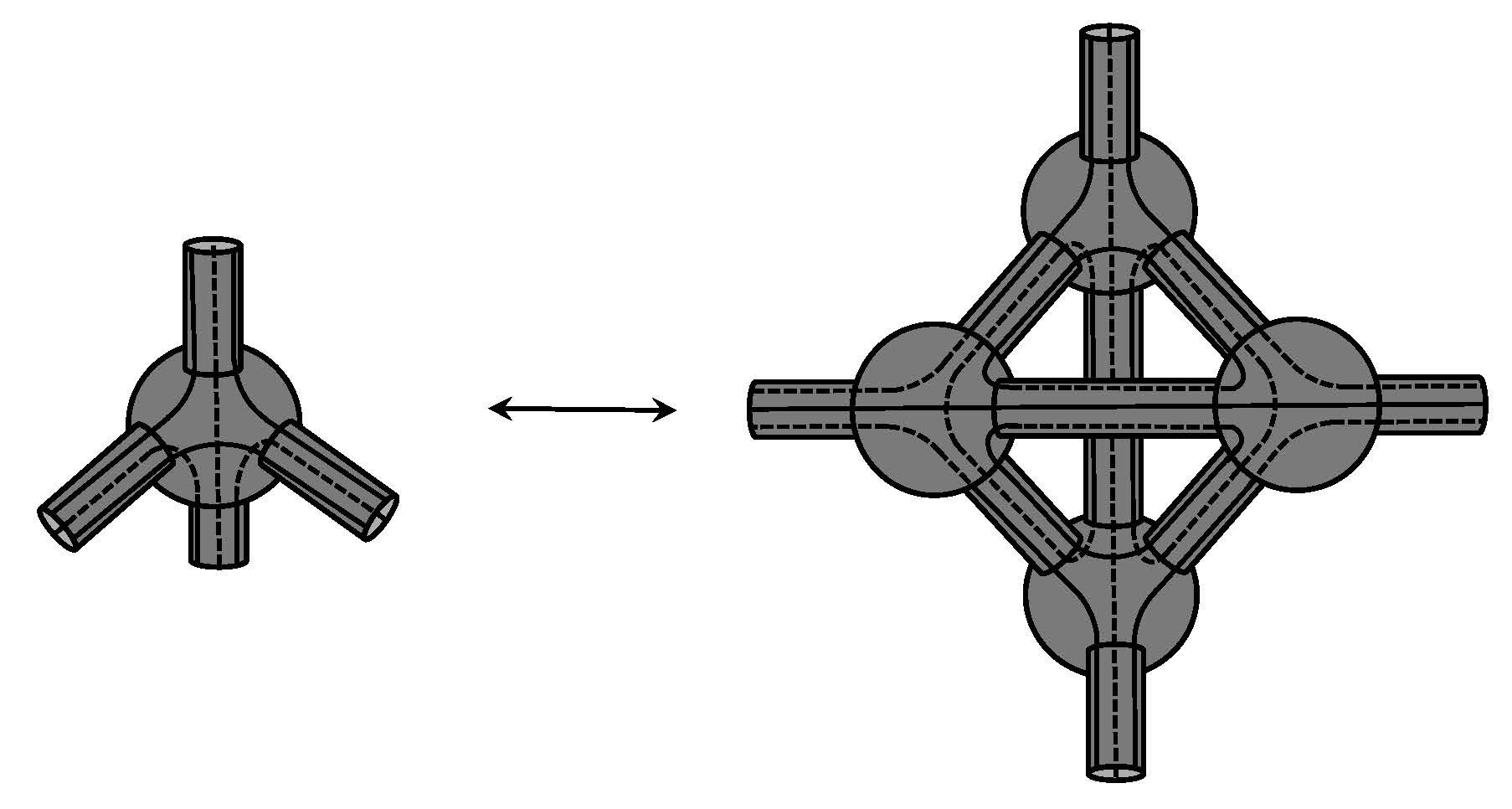}
  \end{center}
\caption{The 1-4 and 4-1 evolution moves} \label{12}
\end{figure}

For $4$ valent networks, the corresponding simplices are tetrahedra. Here the Pachner moves are the $2-3$ and $3-2$ moves (figure \ref{9}), and the $1-4$ and $4-1$ moves (figure \ref{10}). The corresponding evolution moves on braided ribbon networks are then figures \ref{11}, and \ref{12}.

In all of these cases we require that the moves are only allowed if they can be performed smoothly, that is to say that the operation can be performed through smooth deformations of the 2-surface together with point-like changes in genus of the surface which occur at points which are in the interior of the $3$-manifold. This restricts us from performing something like the $1-3$ move in such a manner that the three nodes now encircle some other part of the 2-surface or a feature of the topology of the 3-manifold (such as would occur if the space were a 3 dimensional torus).

\section{Invariants of Braided Ribbon Networks}

We now move on to the next goal of this paper: to demonstrate the existence of invariants of the braided ribbon networks under these evolution moves. To do this we first observe the fact that the evolution moves are defined in such a way that the external edges (and stripes) are unchanged. We will first study trivalent networks and then demonstrate that the invariant for these networks likewise works for $4$-valent BRNs. The invariant that we present for trivalent networks has been presented elsewhere originally in \cite{BilsonThompson:2006yc}, but in a slightly different context.
We consider the set of curves defined by the racing stripes in the manifold which the braided ribbon network is embedded in, which we call the \textit{stripe diagram} of the BRN.

\begin{quote}
\textbf{Theorem} For a BRN generated from a finite graph, the curves which constitute the stripe diagram are only loops.\\
\textbf{Proof} We prove by contradiction. Assuming that there is a curve which is not a loop, this curve can be parameterized and - as the graph is finite - has an origin and a terminus. The racing stripes are formed by composing paths along the edges and the nodes, with the stripe along each edge connecting two nodes and the stripe along each node connecting two edges. We thus find that there are no points which could act as an origin or a terminus, and we thus have a contradiction.
\end{quote}

The stripe diagram is thus a collection of loops embedded in a manifold. We now make an observation: examining the $1-3$ move we see that the operation changes the corresponding stripe diagram by the introduction of a single loop. Also due to the restriction we imposed in section \ref{pachner}, we see that this loop cannot be linked with any other loop, or be a non-trivial element of the fundamental group. From this insight we ask the following: what if we considered the stripe diagram, removing any such loops?

\begin{quote}
We construct the \textit{Reduced Link} of the three valent BRN by removing from the stripe diagram all unlinked and unknotted loops which are isotopic to the identity of the fundamental group of the 3-manifold the BRN is embedded in.
\end{quote}

\begin{figure}[!h]
  \begin{center}
    \includegraphics[scale=0.2]{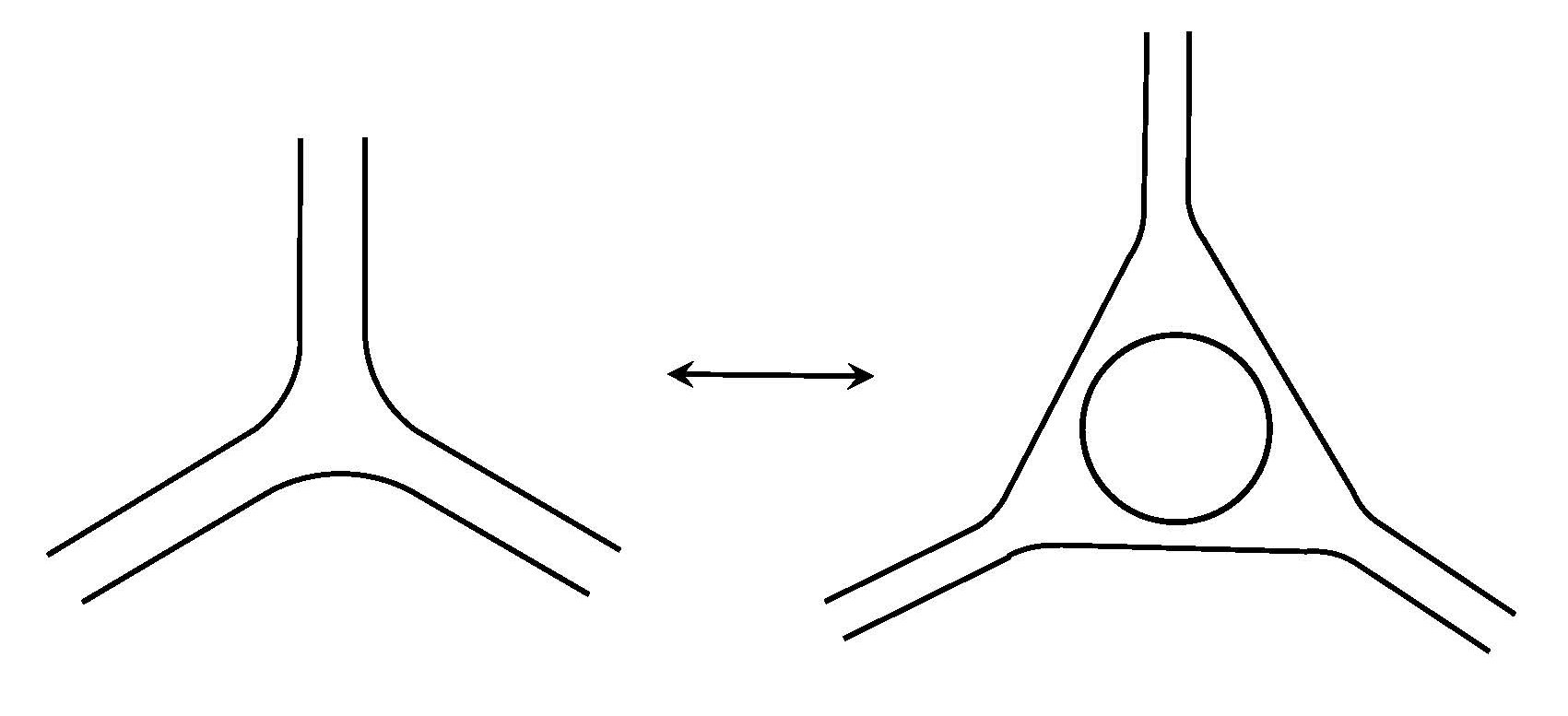}     \includegraphics[scale=0.2]{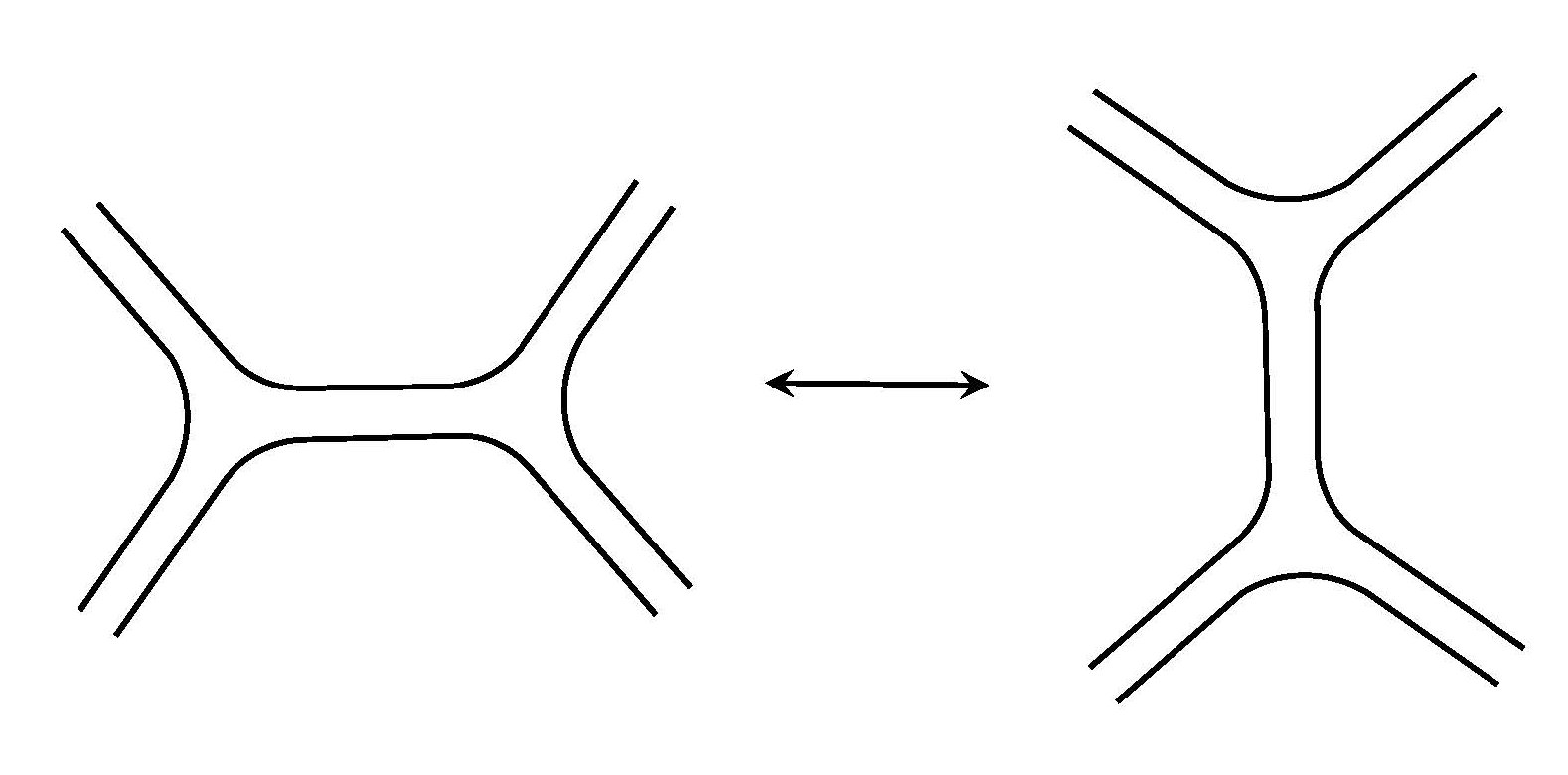}
  \end{center}
\caption{Stripe Diagrams of 3 valent moves} \label{13}
\end{figure}

Now we demonstrate that the reduced link of the three valent BRN is in fact an invariant of the evolution moves (the existence of this proof was originally mentioned in \cite{Hackett:2007dx} for the original picture of trivalent BRNs). To see this, we consider the reduced link of a BRN both before and then after the application of each of the evolution moves. The contributions from the involved nodes is unchanged as is demonstrated by the stripe diagrams in figure \ref{13}, and it is clear from these that the reduced link of a network is preserved under the evolution moves.

\begin{figure}[!h]
  \begin{center}
    \includegraphics[scale=0.2]{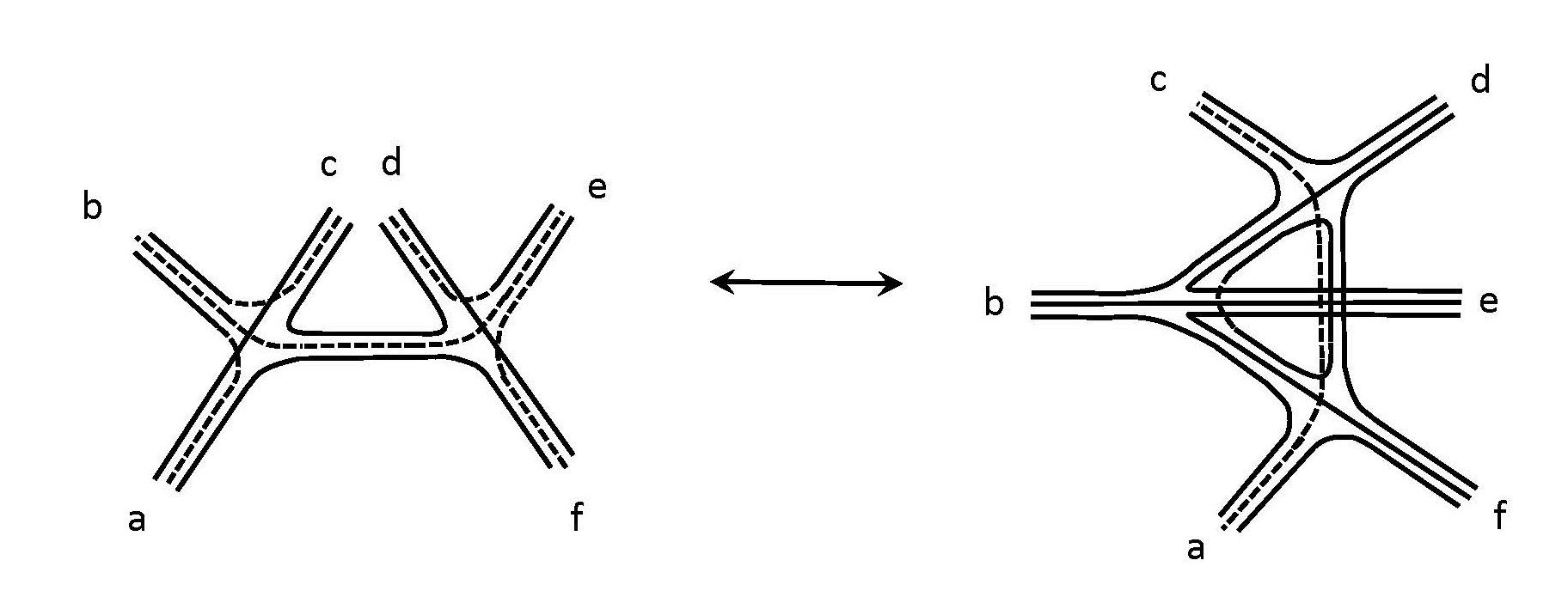}     \includegraphics[scale=0.2]{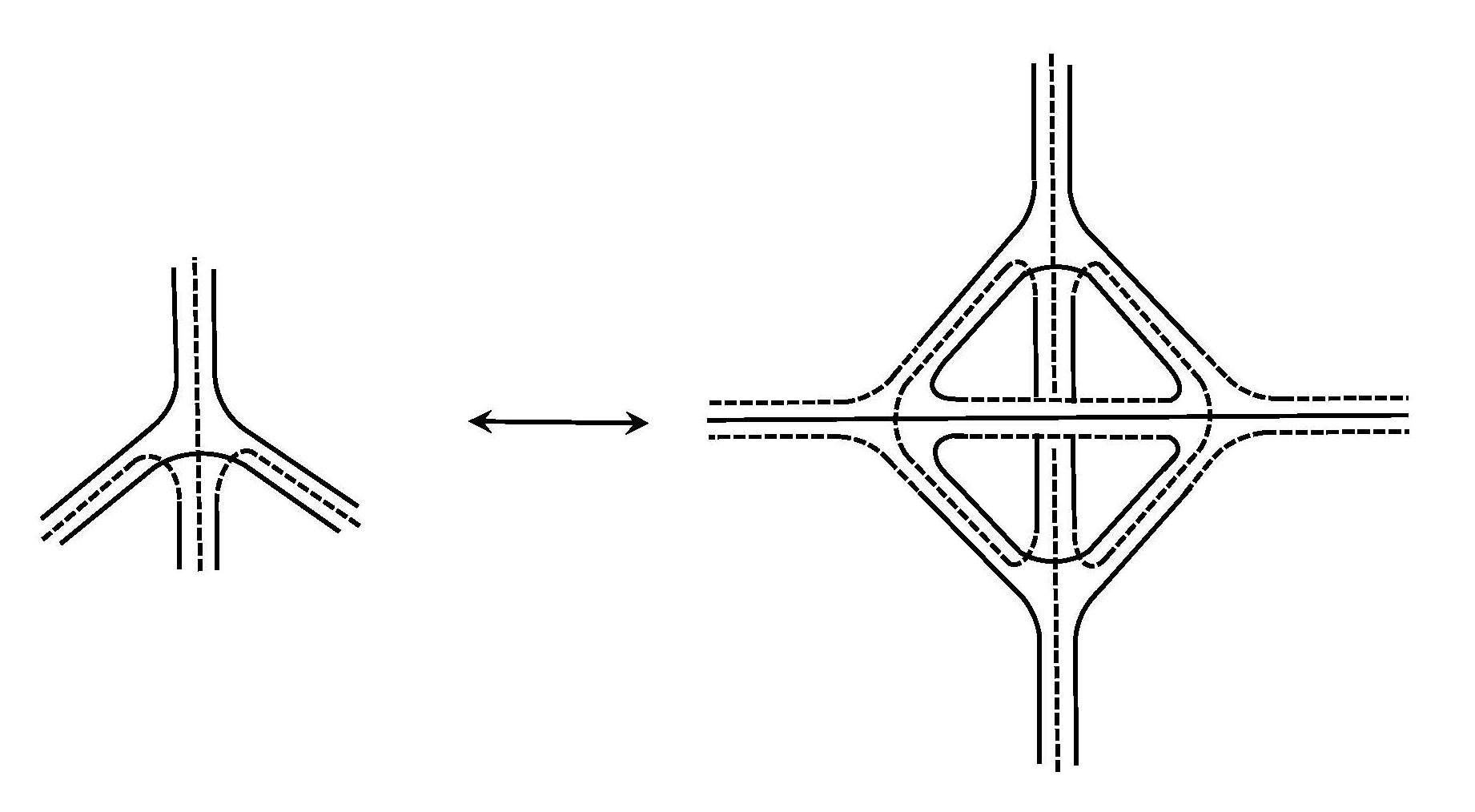}
  \end{center}
\caption{Stripe Diagrams of 4 valent moves} \label{14}
\end{figure}

We now extend the definition of the reduced link to the four valent BRNs. This extension is obvious: there is nothing in the definition of the reduced link that would prevent us from taking it - as is - for 4-valent BRNs as well. Thus all that remains is to demonstrate that the reduced link is also invariant under the 4-valent evolution moves. We do this - as above - through looking at the stripe diagrams in figure \ref{14}.

Looking at these figures, we see that all the stripes connect the same exterior edges as before (this corresponds to the exterior faces and edges of the glued tetrahedra being unchanged by the pachner move). It only remains to examine whether any linking, knotting or non-trivially contractible curves have been introduced. We demonstrate this by going step by step through the deformations that give rise to the evolution moves (we will only proceed in a single direction, as the processes are invertible).

The $1-4$ move is achieved, beginning with a single node, by introducing four closed loops on the surface of the sphere one in each of the four regions bounded by the racing stripes - these loops each correspond to the introduction of an edge in the dual pachner move. From here we separate the sphere into four parts (this process is allowed under the caveat of point-like changes in the genus in the interior of the manifold), leaving the only aesthetic deformations remaining (straightening out the connections between the sphere into tubes etc). In this process, the only changes to the stripe space were the introduction of the four closed loops on the surface of the sphere. By definition these cannot have any knotting or linking, and must be in the trivial element of the fundamental group. These therefore are removed in the map from the stripe space to the reduced link, and so leave the reduced link unchanged.

The $2-3$ move begins with $2$ nodes connected by a single edge. To make the steps clear we will label the edges as in the diagram, and though the steps we take will specify particular edges they aren't unique and so can be made general through relabeling. This interior edge is expanded to combine it and the two spheres into a single sphere. We now introduce a single closed loop corresponding to the introduction of the single edge in the dual pachner move. This loop is placed in the region interior to the stripes which connect $f$ to $d$, $d$ to $e$, and $e$ to $f$ (or equivalently $a$ to $c$, $c$ to $b$, and $b$ to $a$). Splitting the sphere into a torus through the center of this loop, we then regroup the edges into pairs $c$ with $d$, $b$ with $e$ and $a$ with $f$. For aesthetics we constrict the region of the sphere between the pairs of edges into tubes, making our three nodes. We can see here that the only change to the stripe space was the introduction of a single trivial loop. This change does not alter the reduced link. Completing the proof of the invariance of the reduced link.

\subsection{Why do we call these ribbons anyway?}

We will take this opportunity to explain why these graphs made from tubes and spheres are referred to as braided ribbon graphs by demonstrating the connection to the graphs of \cite{BilsonThompson:2006yc}, and as promised show that the definition of a reduced link coincides in the two pictures.

\begin{figure}[!h]
  \begin{center}
    \includegraphics[scale=0.2]{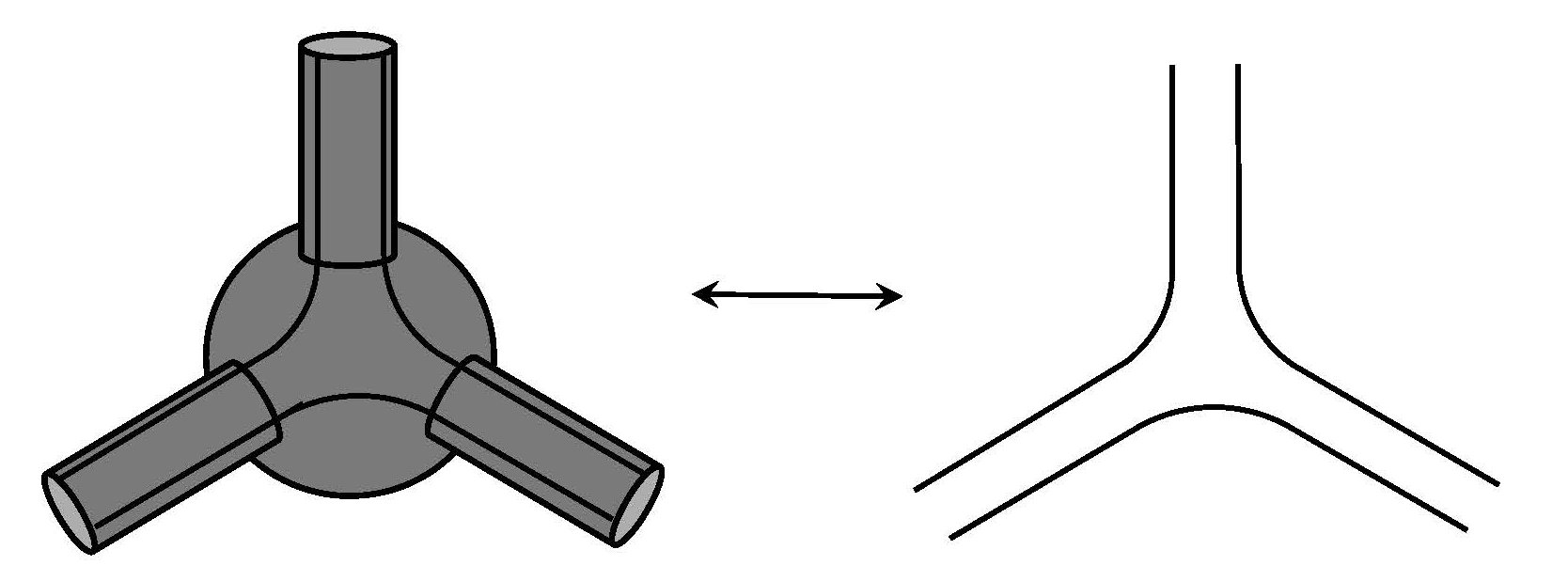}
  \end{center}
\caption{From BRN to trinions} \label{15}
\end{figure}

In \cite{BilsonThompson:2006yc} trivalent braided ribbon networks are defined as two dimensional surfaces constructed by taking the union of trinions (2-surfaces with three distinct `legs' along which they can be connected to one another) as shown in figure \ref{15}. Each trinion defines a node, and each of the legs of the trinions is an edge. Like the edges of braided ribbon networks the legs which connect the trinions to one another can be braided, twisted or knotted. These ribbon graphs are considered up to equivalence class defined by smooth deformations of the resulting surfaces.

We construct from one of these ribbon graphs a braided ribbon network as we've defined in section \ref{BRNS} as follows: for each node of the network we consider a closed ball in the embedding space which has the node on its surface but which has an empty intersection with the rest of the ribbon graph. These spheres then define the nodes of the braided ribbon network. The edges of the braided ribbon networks are then defined by similarly constructing tubes between these spheres so that the boundaries of the edges of the ribbon graph coincide with the boundary of the tubes. The boundaries of the surface of the ribbon graph then become the racing stripes of the braided ribbon network.

Likewise we can construct a ribbon graph from a 3-valent braided ribbon network by making the following observation: at each node the racing stripes divide the sphere into two parts, likewise along each edge the tube is divided in two by the racing stripes. We can consistently choose one side or the other and identify this as the surface of a ribbon graph (alternatively we can think of `squishing' the two halves together into a single surface, in a sort defining one side to be the `front' and the other the `back').

From this identification we can see the correspondence between the reduced link defined on these two equivalent types of graph. In \cite{BilsonThompson:2006yc} the reduced link is defined by considering the boundary of the 2-surface independently and removing unlinked and unknotted curves. Here we've seen that the boundary of the 2-surface is equivalent to the racing stripes of the BRN, and thus that this operation is equivalent to the reduced link we've defined.

\section{Conclusions}

We have provided a unified framework for studying braided ribbon networks which incorporates all previous results. Using this framework we have generalized the reduced link and have also shown that the reduced link is an invariant of the evolution moves of the braided ribbon network. The promise in this result is that it should allow us to proceed to study these networks from a consistent basis and allow for results in one valence to be applied to another.

Having a consistent basis also us to more easily connect these results to other mathematical objects. It may also be possible to relate these results to the other varieties of networks laid out in \cite{Markopoulou:2008be} and to thus use the concept of a reduced link for some of them as well. The ability to bring these tools and the related work in attempting to find particle-like excitations into other mathematical frameworks would be an extremely promising step forward in attempts to understand background independent approaches to quantum gravity and their potential unification with particle physics.

\section*{Acknowledgements}

The author is indebted to his Advisor Lee Smolin, for his
discussion and critical comments, additionally he - along with Yidun Wan - deserve thanks for providing the motivation for this work in their original work in the 4-valent formalism. Thanks are also extended to the group at Marseille for hosting the author during the time this work was cemented and for their critical discussion of the work. Particular thanks are due to Adam Ramer for his providing the figures contained within.  This work was supported by the Government of Canada through NSERC's CGS-MSFSS program and NSERC's CGS program. Research at Perimeter Institute for
Theoretical Physics is supported in part by the Government of Canada
through NSERC and by the Province of Ontario through MRI.


\begin{thebibliography}{99}
\bibitem{BilsonThompson:2005bz}
  S.~O.~Bilson-Thompson,
  ``A Topological model of composite preons,''
  Submitted to: Phys.Lett.B.
  [hep-ph/0503213].

\bibitem{BilsonThompson:2006yc}
  S.~O.~Bilson-Thompson, F.~Markopoulou and L.~Smolin,
  ``Quantum gravity and the standard model,''
  Class.\ Quant.\ Grav.\  {\bf 24}, 3975 (2007)
  [arXiv:hep-th/0603022].

\bibitem{Hackett:2007dx}
  J.~Hackett,
  ``Locality and translations in braided ribbon networks,''
  Class.\ Quant.\ Grav.\  {\bf 24 } (2007)  5757-5766.
  [hep-th/0702198 [HEP-TH]].
\bibitem{Wan:2007nf}
  Y.~Wan,
  ``On Braid Excitations in Quantum Gravity,''
  [arXiv:0710.1312 [hep-th]].

\bibitem{Smolin:2007sn}
  L.~Smolin, Y.~Wan,
  ``Propagation and interaction of chiral states in quantum gravity,''
  Nucl.\ Phys.\  {\bf B796}, 331-359 (2008).
  [arXiv:0710.1548 [hep-th]]

\bibitem{Hackett:2008ie}
  J.~Hackett and Y.~Wan,
  ``Infinite Degeneracy of States in Quantum Gravity,''
  arXiv:0811.2161 [hep-th].


\bibitem{BilsonThompson:2008ex}
  S.~Bilson-Thompson, J.~Hackett, L.~Kauffman, L.~Smolin,
  ``Particle Identifications from Symmetries of Braided Ribbon Network Invariants,''
  [arXiv:0804.0037 [hep-th]].

\bibitem{Hackett:2008tt}
  J.~Hackett, Y.~Wan,
  ``Conserved Quantities for Interacting Four Valent Braids in Quantum Gravity,''
  Class.\ Quant.\ Grav.\  {\bf 26}, 125008 (2009).
  [arXiv:0803.3203 [hep-th]].




\bibitem{He:2008is}
  S.~He, Y.~Wan,
  ``Conserved Quantities and the Algebra of Braid Excitations in Quantum Gravity,''
  Nucl.\ Phys.\  {\bf B804}, 286-306 (2008).
  [arXiv:0805.0453 [hep-th]].

\bibitem{He:2008jc}
  S.~He, Y.~Wan,
  ``C, P, and T of Braid Excitations in Quantum Gravity,''
  Nucl.\ Phys.\  {\bf B805}, 1-23 (2008).
  [arXiv:0805.1265 [hep-th]].

\bibitem{Wan:2008qs}
  Y.~Wan,
  ``Effective Theory of Braid Excitations of Quantum Geometry in terms of Feynman Diagrams,''
  Nucl.\ Phys.\  {\bf B814}, 1-20 (2009).
  [arXiv:0809.4464 [hep-th]].


\bibitem{BilsonThompson:2009fh}
  S.~Bilson-Thompson, J.~Hackett and L.~H.~Kauffman,
  ``Particle Topology, Braids, and Braided Belts,''
  J.\ Math.\ Phys.\  {\bf 50}, 113505 (2009)
  [arXiv:0903.1376 [math.AT]].

\bibitem{Markopoulou:1997hu}
  F.~Markopoulou and L.~Smolin,
  ``Quantum geometry with intrinsic local causality,''
  Phys.\ Rev.\  D {\bf 58}, 084032 (1998)
  [arXiv:gr-qc/9712067].

\bibitem{Smolin:2002sz}
  L.~Smolin,
  ``Quantum gravity with a positive cosmological constant,''

  [hep-th/0209079].
\bibitem{Crane:1991ke}
  L.~Crane,
  ``Conformal field theory, spin geometry, and quantum gravity,''
  Phys.\ Lett.\  {\bf B259}, 243-248 (1991).

\bibitem{Crane:1991jv}
  L.~Crane,
  ``2-d physics and 3-d topology,''
  Commun.\ Math.\ Phys.\  {\bf 135}, 615-640 (1991).

\bibitem{kk}
K.~ Kuratowski, ``Sur le problème des courbes gauches en topologie,'' Fund. Math., {\bf15}, 271-283, 1930.

\bibitem{Markopoulou:2008be}
  F.~Markopoulou, I.~Premont-Schwarz,
  ``Conserved Topological Defects in Non-Embedded Graphs in Quantum Gravity,''
  Class.\ Quant.\ Grav.\  {\bf 25}, 205015 (2008).
  [arXiv:0805.3175 [gr-qc]].


\end{thebibliography}
\end{document}